# IMPLICATIONS OF QUANTUM COMPUTING FOR ARTIFICIAL INTELLIGENCE ALIGNMENT RESEARCH

by Jaime Sevilla[1] and Pablo Moreno[2]

**ABSTRACT:** We explain the key features of quantum computing via three heuristics and apply them to argue that a deep understanding of quantum computing is unlikely to be helpful to address current bottlenecks in Artificial Intelligence Alignment.

Our argument relies on the claims that Quantum Computing leads to compute overhang instead of algorithmic overhang, and that the difficulties associated with the measurement of quantum states do not invalidate any major assumptions of current Artificial Intelligence Alignment research agendas.

We also discuss tripwiring, adversarial blinding, informed oversight and side effects as possible exceptions.

**EPISTEMIC STATUS:** Exploratory, we could have overlooked key considerations.

# Introduction

Quantum Computing (QC) is a disruptive technology that may not be too far ahead in the horizon. Small proof-of-concept quantum computers have already been built [1] and major obstacles to large-scale quantum computing are being heavily researched [2].

Among its potential uses, QC will allow breaking classical cryptographic codes, simulate large quantum systems and faster search and optimization [3]. This last use case is of particular interest to Artificial Intelligence (AI) Strategy. In particular, variants of the Grover algorithm can be exploited to gain a quadratic speedup in search problems, and some recent Quantum Machine Learning (QML) developments have led to exponential gains in certain Machine Learning tasks [4] (though with important caveats which may invalidate their practical use [5]).

These ideas have the potential to exert a transformative effect on research in AI (as noted in [6], for example). Furthermore the technical aspects of QC, which put some physical limits on the observation of the inner workings of a quantum machine and hinder the verification of quantum computations [7], may pose an additional challenge for AI Alignment concerns.

In this short article we introduce a heuristic model of quantum computing that captures the most relevant characteristics of QC for technical AI Alignment research.

---

[1] jaime.sevillamolina@philosophy.ox.ac.uk
[2] pabloamo@ucm.es

We then apply our model to abstractly answer in which areas we expect knowledge of QC might be relevant, and discuss four specific avenues of current research where it might come into play: tripwiring, adversarial blinding, informed oversight and avoiding side effects.

# A model of Quantum Computing for AI Alignment

Here we give a very short and simplified introduction to QC for AI Alignment Researchers. For a longer introduction to QC we recommend *Quantum Computing for the Very Curious* [8]. If you're already familiar with QC you may want to check our technical refresher in the footnotes in appendix A.

We introduce **three heuristics,** which to the best of our knowledge capture all relevant aspects of QC for AI Alignment concerns:

1. **Quantum speedup** - quantum computers are, at most, as powerful as classical computers we allow to run for an exponential amount of time .

   QC usually just gets you quadratic advantages (ie Grover database search [9], Quantum Walks [10]...), although in some cases they are exponential with respect to the best known classical algorithms (eg. Shor [11], Hamiltonian simulation [12], HHL [13]...).

   Technically, the class of problems efficiently solvable with a probabilistic classical computer (BPP) is a subset of the class of problems efficiently solvable by a quantum computer (BQP), and that one is itself a subset of the problems solvable in exponential time by a classical deterministic computer (EXP). This means that QC are at least as fast as classical computers, but no more than exponentially faster [14, 15].

2. **Quantum obfuscation** - there is no efficient way of reading the state of a quantum computer while it is operating.

   Quantum operations cannot copy quantum states ("No-cloning theorem") [16, 17] and performing a partial or total measure of a quantum state will collapse that part of the state, resulting in loss of information.

   To recover information from a quantum state one has several inefficient options: performing the inner product of the state with another vector (usually through a procedure called the swap test [18]), perform many measurements on many identically prepared states to do statistics on the entries (tomography [19]), or using amplitude estimation [20] to estimate a single amplitude.

   The former procedure depends on the precision quadratically and destroys the state, albeit it is independent of the dimension of the vector. The latter two require a number of repetitions at least linear with respect to the dimension of the vector, which grows exponentially with the number of qubits.

3. **Quantum isolation** - a quantum computer cannot interact with the classical world without its state becoming at least partially classic.

   In other words, if a quantum computer creates a side-channel to the outside world during their quantum computation it destroys its coherence and randomizes state according to well defined rules.

   This is directly derived from the postulates of Quantum Mechanics, and in particular the collapse of the wave function when it interacts with the outside world.

In the following two sections we look at how this model can be applied to gather insight on the phases of research in AI Alignment and kinds of alignment strategies where QC may or may not be relevant.

## Bottlenecks in Artificial Intelligence Alignment research

In this section we introduce a simplified way of thinking about the different phases of research through which we expect the field of AI Alignment to go, and reason about the relevance of quantum computing during each of these phases.

Looking at some landmark achievements in computer science, it seems that most research begins with working on the formalization of a problem, which then is followed by a period where researchers try to find solutions to the problem, at first just theoretical, then inefficient and finally practical implementations (see for example the history of chess playing, from Shannon's seminal paper in 1950 to the Deeper Blue vs Kasparov match in 1997 [21]).

We expect research in AI Alignment to develop in a similar fashion, and while we have some formalized frameworks to handle some subsets of the problem (see for example IRL [22]), there is no agreed upon formalization that captures the essence of the whole alignment problem.

On the other hand, theoretical proposals for QC applications are mostly concerned with speeding up classical algorithms, sometimes with notable improvements (see for example Shor's algorithm for factorization [11]), and in some rare cases it has inspired the creation of novel algorithmic strategies [23]. In no case that we know of has QC lead to a formalization insight of the kind that we believe AI Alignment is bottlenecked on.

That is, QC has so far only helped find efficient solutions to problems that were already properly formalized, while we believe that the most significant problems in AI Alignment have not yet matured into proper formalizations.

This observation serves as an empirical verification of the quantum speedup heuristic, that instructs us to think about quantum computing as a black box accelerator rather than a novel approach to algorithmic design, and thus we should not expect formalization insights to come from QC. In other words, QC may lead to what would be equivalent to compute overhang, but not lead to significant insight overhang.

We conclude that while QC may help in a later phase of AI Alignment research with making safe AI algorithms practical and competitive, it is very unlikely that it will lead to novel theoretical insights that fundamentally change how we think about AI Alignment.

As a side note, the same reasoning applies to AI capabilities research; QC is unlikely to lead to new formal insights on that field. However, the quantum speedup may enable the practical use of algorithms which were previously considered inefficient. This is concerning to the extent that we expect compute overhang to lead to more opaque and/or less safe algorithms.

## Alignment Strategies: incentive design versus active oversight

In this section we introduce a distinction between two main broad complementary strategies for achieving AI Alignment: incentive design and active oversight, and reason about how QC may interact with them.

By **incentive design** we mean static strategies, where the design of an agent is verified to have certain safety properties that incentivize the agent to pursue desirable goals.

By **active oversight** we refer to dynamic strategies, where an agent, which may or may not be safe, is monitored, and if certain metrics indicate unsafeness, an intervention is made to safely interrupt or modify the agent.

We believe that a complete solution to the AI Alignment problem will include both elements of incentive design and active oversight.

Since we can treat QC as a black box accelerator according to the quantum speedup heuristic, we can see that QC does not present any additional challenges for incentive design. In other words, since designing the right incentives for the behaviour of the agent is independent of the hardware and implementation efficiency, QC is unlikely to help with this endeavour.

Instead of concerning themselves with the intricacies of QC, AI Alignment researchers may just treat QC as a black box and reason about classical equivalents instead. This is analogous to how AI Safety researchers do not need to think about whether an agent will be implemented over GPUs or serial processors and only focus on the formalisms.

Those alternatives may be exponentially slower; however, if the safety of our agent design relies on the assumption that the agent will not be able to acquire exponentially more resources for computation we cannot really deem it safe, making this a moot point.

One possible counter-argument to our line of reasoning above is that quantum computing may influence which AI designs end up being successful, and knowledge about the AI designs that are going to be more popular in the future can allow us to make extra assumptions to facilitate incentive design. We leave as an open question whether we can reliable predict features about QC-powered agents that will allow us to make these kind of simplifying assumptions.

Moving on from incentive design, we point out that QC may pose an additional challenge to active oversight due to quantum obfuscation, as it might interfere with measures of the internal state of a QC. In the next section we explore some AI Alignment research areas where this fact comes into play.

# Review of AI Alignment research agendas

We have reviewed the AI Alignment agendas of *Concrete Problems in AI Safety* [24], MIRI's research agendas [25] and *Iterated Distillation and Amplification* [26].

Most of their work focuses on incentive design as the de facto route to a solution. As such, QC does not affect the main assumptions and considered solutions. However, we have been able to identify four areas of active oversight where QC may be a relevant consideration: tripwiring, adversarial blinding, informed oversight and avoiding side effects.

We discuss each of these areas in depth.

## Tripwiring

**Tripwiring** is a class of mechanisms that alert of, or directly prevent attempts of an agent to hack its reward function [24]. If a particular tripwire design relies on access to the internal state of the agent, quantum obfuscation may impede its effective implementation.

For example, in the Transparency section of *Techniques for optimizing worst-case performance* [27], Paul Christiano writes:

> "While training a model, we are free to look inside and 'see what it's thinking'. If we could actually understand those thoughts, we might be able to more easily predict under what conditions it would fail malignantly."

Unfortunately, due to quantum obfuscation, this seems rather complicated if the agent has quantum capabilities. You will just be able to efficiently read the input/output of the used quantum subroutines, but not its inner operation.

Nevertheless, there might be easier ways for a misaligned agent to obfuscate what it is doing, such as classically encoding its hidden plans in ways that an overseer will not understand (for example, it can encode its state with quantum-resistant cryptography [28]). However some early work points in the direction that quantum-based obfuscation may be significantly more powerful than classical alternatives [29].

Another consideration is that quantum obfuscation offers a more plausible excuse for obfuscation than classical cryptography. For a classical agent that doesn't need to encrypt information during its expected workflow, any encryption attempt might indicate misalignment and trigger a shutdown. If, in contrast, the QC agent hides important information in a quantum superposition, this might be misaligned behavior or just a consequence of wanting to exploit the quantum speedup.

Finally, one could imagine some kind of quantum tripwire that monitors the quantum computer from within the superposition. For example, a quantum bit that flags whether the tripwire has been activated. Since quantum computing allows for conditional operations, one could imagine freezing the state if such state has been reached in the superposition, only allowing it to proceed while it has not. Then one could perform amplitude estimation to check that the flag has not been activated in any branch of the computation, and otherwise kill it.

## Adversarial blinding

**Adversarial blinding** consists of using adversarial techniques to blind the agent with respect to some variables. This technique could be used to make it difficult for the agent to understand how its reward is generated and therefore how it can be hacked [24].

In the case where there is an asymmetry of resources and the overseer has access to quantum computers but the agent does not, we know that there are distributions that are just not efficiently computable classically (will need exponential resources), but are efficiently computable quantically. This suggests the possibility of using one of those distributions to generate a difficult to hack reward.

## Informed oversight

In the **informed oversight** scenario, the verifier checks that the behavior of an agent is safe, and the requirements on the verifier do not grow too much as the agent becomes more and more powerful [30].

If there is an asymmetry in resources where the verifier has access to QC while the agent does not, this may be exploited using quantum speedup to allow the verifier to do certain checks in polynomial time for operations that will take the agent exponential time to check themselves.

In particular, quantum speedup may help with the amplification step of IDA, where we give an agent access to QC to design and verify a stronger successor, whose resources will be constrained as in not having access to QC themselves during the verification step.

It is however unclear whether this scheme of only giving access to the agent to QC after it has been verified is possible, and whether we can reliably restrain the agent from accessing QC while it is being verified.

## Side effects and AI boxing

We would like to be able to reliably detect and prevent AI systems from interacting with the world in unexpected ways (this is referred to in the literature as **avoiding side effects** [24]).

Per the quantum isolation heuristic, when a process implemented in a quantum computer tries to affect the outside world it has to necessarily partially collapse part of the state (unless the state is already classical).

One may think of taking advantage of this phenomena to monitor whether a quantum agent has interacted with the outside world. There are however some associated difficulties.

Firstly, and most importantly, we cannot distinguish a partial collapse from a constructive interference when the amplitude of the subset of qubits we are interested in is 1. That is, it may be the case that part of the state is classical and we cannot effectively tell whether this is due to collapse or because all states in superposition agree on the state of a particular subset of qubits.

Secondly, to tell whether the process has (partially) collapsed we need to measure it, which will collapse it in turn, and even worse, we will not be able to know whether the state had collapsed before or because of our measurement. Thus we would need to resort to one of the inefficient ways of measuring the quantum state, like amplitude estimation.

All this means it does not seem possible to strongly and efficiently infer interaction with the outside world through learning that part of the state of the agent's mind is classical.

Relatedly, **AI boxing** refers to restraintment strategies where an AI system is isolated from the environment to prevent unexpected side effects [31].

The quantum isolation heuristic may mislead us into thinking that boxing may be easier to achieve in QC. However this is not the case; as long as a process inside a QC is willing to partially collapse part of its state it can escape its quantum box. In fact, one of the most active areas of current research in QC is investigating how to isolate the quantum system so it does not collapse unexpectedly.

So, in conclusion, one may think of quantum computers as oracles, but that does not mean that they are naturally boxed already.

# Conclusion and open questions

We have presented a simplified model of quantum computing for AI Alignment, summarized in three heuristics: quantum speedup (QC allows a quadratic and sometimes exponential speedup), quantum obfuscation (qubits cannot be read without destroying the computation) and quantum isolation (interacting with the classical world destroys any superposition).

Our main conclusion is that we do not expect Quantum Computing to be relevant to the current technical AI Alignment research agendas.

In particular, we do not expect QC techniques to be relevant for safety until we are actually interested in making algorithms and protocol as efficient as possible.

Furthermore, we expect QC to not be relevant for general incentive design purposes. However, we have also listed some current research problems related to active oversight where it might be helpful to have our simplified model of quantum computing in mind. In any case we do not expect these issues to have high relevance right now, as most of the current work in current AI Alignment falls under incentive design strategies rather than active oversight.

Some further questions that came up during our research:

- Can we expect humanity to develop reliable QC before AGI? How do the developments in each field interact with one another? Will quantum machine learning significantly increase AI capabilities? Since one of the most straightforward and promising applications of QC is material science, should we expect QC to lead to further improvements in non-QC computing technology? How does that affect AI Capabilities?

- In the case where QC is relevant for the design of advanced AIs, can we expect to have an actual quantum agent in the future, or will it just be a classical agent with access to quantum subroutines, in a Comprehensive AI Services fashion [32]?

- How does QC affect AI Governance? How easy is to deploy powerful QC-powered AI systems? How easily can we monitor the labs that have QC capabilities? If QC is relevant to the development of AGI, can the fact that QC capabilities are more concentrated affect the dynamics of development of AGI?

- How does the quantum speedup possibility affect AI design? Can developments in QC lead to opaque AI, as the compute improvements allow raw search design to be used instead of design from first principles? Relatedly, what AI designs would be particularly favored by QC technology versus raw improvements in compute?

- Is quantum obfuscation the most efficient obfuscation strategy a misaligned agent can have access to? Can classical cryptography be used to obfuscate information?

- Is asymmetry of resources a reasonable assumption to make in verifier / agent scenarios? How can asymmetry of QC resources be exploited for safety purposes?

- How could we design quantum tripwires? What are their strengths and limitations?

- How would we go about implementing an adversarial blinding scheme based on quantum distribution sampling?

---

Article by Jaime Sevilla (FHI summer research fellow) and Pablo Moreno (Quantum Computing PhD student at Complutense University of Madrid under an FPU grant).

We want to thank Linh Chi Nguyen, Adrian Hutter, Anders Sandberg, Max Daniel, Richard Möhn and Daniel Eth for incredibly useful feedback, editing and discussion.

Daniel Eth contributed directly to the collection of open questions. Anders Sandberg pointed us to the isolation heuristic and its possible implications. Adrian Hutter prevented us from making a wrong claim on the complexity bounds of BQP.

## Appendix A: Speed technical introduction to Quantum Computing

Quantum states are complex unitary vectors. A basis vector is just a classical state, whereas any other is called a superposition (a linear combination of basis states) [33]. Quantum Computing is based on unitary transformations of these quantum states. Non-unitary dynamics can be introduced via measurements: a measurement projects the quantum state into a basis state (classical state) with a probability equal to the square of the amplitude of that state (the coefficient in the linear combination).

## Bibliography

[1] Córcoles, A. D., et al. «Demonstration of a Quantum Error Detection Code Using a Square Lattice of Four Superconducting Qubits». Nature Communications, vol. 6, n.o 1, noviembre de 2015, p. 6979. DOI.org (Crossref), doi:10.1038/ncomms7979.

[2] Almudever, C. G., et al. «The engineering challenges in quantum computing». Design, Automation & Test in Europe Conference & Exhibition (DATE), 2017, IEEE, 2017, pp. 836-45. DOI.org (Crossref), doi:10.23919/DATE.2017.7927104.

[3] de Wolf, Ronald. «The Potential Impact of Quantum Computers on Society». Ethics and Information Technology, vol. 19, n.o 4, 2017, pp. 271-76. DOI.org (Crossref), doi:10.1007/s10676-017-9439-z.

[4] Lloyd, Seth, y Christian Weedbrook. «Quantum Generative Adversarial Learning». Physical Review Letters, vol. 121, n.o 4, 2018, p. 040502. DOI.org (Crossref), doi:10.1103/PhysRevLett.121.040502.

[5] Aaronson, Scott. «Quantum Machine Learning Algorithms: Read the Fine Print». https://www.scottaaronson.com/papers/qml.pdf.

[6] Dafoe, Allan. AI Governance: A Research Agenda. Future of Humanity Institute, University of Oxford, https://www.fhi.ox.ac.uk/wp-content/uploads/GovAIAgenda.pdf.

[7] Mahadev, Urmila. «Classical Verification of Quantum Computations». 2018 IEEE 59th Annual Symposium on Foundations of Computer Science (FOCS), IEEE, 2018, pp. 259-67. DOI.org (Crossref), doi:10.1109/FOCS.2018.00033.

[8] Matuschak, Andy, and Nielsen, Michael. «Quantum Computing for the very curious». https://quantum.country/qcvc.

[9] Grover, Lov K. «A Fast Quantum Mechanical Algorithm for Database Search». *Proceedings of the Twenty-Eighth Annual ACM Symposium on Theory of Computing  - STOC '96*, ACM Press, 1996, pp. 212-19. *DOI.org (Crossref)*, doi:10.1145/237814.237866.

[10] Szegedy, M. «Quantum Speed-Up of Markov Chain Based Algorithms». 45th Annual IEEE Symposium on Foundations of Computer Science, IEEE, 2004, pp. 32-41. DOI.org (Crossref), doi:10.1109/FOCS.2004.53.

[11] Shor, Peter W. «Polynomial-Time Algorithms for Prime Factorization and Discrete Logarithms on a Quantum Computer». SIAM Journal on Computing, vol. 26, n.o 5, octubre de 1997, pp. 1484-509. DOI.org (Crossref), doi:10.1137/S0097539795293172.

[12] Berry, Dominic W., et al. «Hamiltonian Simulation with Nearly Optimal Dependence on all Parameters». 2015 IEEE 56th Annual Symposium on Foundations of Computer Science, IEEE, 2015, pp. 792-809. DOI.org (Crossref), doi:10.1109/FOCS.2015.54.

[13] Harrow, Aram W., et al. «Quantum Algorithm for Linear Systems of Equations». Physical Review Letters, vol. 103, n.o 15, octubre de 2009, p. 150502. DOI.org (Crossref), doi:10.1103/PhysRevLett.103.150502.

[14] Aaronson, Scott. «BQP and the Polynomial Hierarchy». Proceedings of the 42nd ACM Symposium on Theory of Computing - STOC '10, ACM Press, 2010, p. 141. DOI.org (Crossref), doi:10.1145/1806689.1806711.

[15] Petting Zoo - Complexity Zoo. https://complexityzoo.uwaterloo.ca/Petting_Zoo.

[16] Wootters, W. K., y W. H. Zurek. «A Single Quantum Cannot Be Cloned». Nature, vol. 299, n.o 5886, octubre de 1982, pp. 802-03. DOI.org (Crossref), doi:10.1038/299802a0.

[17] Scarani, Valerio, et al. «Quantum Cloning». Reviews of Modern Physics, vol. 77, n.o 4, noviembre de 2005, pp. 1225-56. DOI.org (Crossref), doi:10.1103/RevModPhys.77.1225.

[18] Schuld, Maria and Petruccione, Francesco. «Supervised learning with quantum computers». Springer Berlin Heidelberg, 2018.

[19] Cramer, Marcus, et al. «Efficient Quantum State Tomography». Nature Communications, vol. 1, n.o 1, 2010, p. 149. DOI.org (Crossref), doi:10.1038/ncomms1147.

[20] Brassard, Gilles, et al. «Quantum Amplitude Amplification and Estimation». Contemporary Mathematics, editado por Samuel J. Lomonaco y Howard E. Brandt, vol. 305, American Mathematical Society, 2002, pp. 53-74. DOI.org (Crossref), doi:10.1090/conm/305/05215.

[21] «CS221». Deep Blue, https://stanford.edu/~cpiech/cs221/apps/deepBlue.html.

[22] Ng, Andrew Y., and Stuart J. Russell. «Algorithms for inverse reinforcement learning.» Icml. Vol. 1. 2000.

[23] Tang, Ewin. «A Quantum-Inspired Classical Algorithm for Recommendation Systems». Proceedings of the 51st Annual ACM SIGACT Symposium on Theory of Computing - STOC 2019, ACM Press, 2019, pp. 217-28. DOI.org (Crossref), doi:10.1145/3313276.3316310.

[24] Amodei, Dario, et al. «Concrete problems in AI safety.» arXiv preprint arXiv:1606.06565 (2016).

[25] Demski, A., & Garrabrant, S.. «Embedded agency. arXiv preprint arXiv:1902.09469. (2019)

[26] Cotra, Ajeya. «Iterated Distillation and Amplification». Medium, 29 de abril de 2018, https://ai-alignment.com/iterated-distillation-and-amplification-157debfd1616.

[27] Christiano, Paul. «Techniques for Optimizing Worst-Case Performance». Medium, 2018, https://ai-alignment.com/techniques-for-optimizing-worst-case-performance-39eafec74b99.

[28] Lily Chen, et al. «Report on Post-Quantum Cryptography». National Institute of Standards and Technology, https://nvlpubs.nist.gov/nistpubs/ir/2016/nist.ir.8105.pdf

[29] Alagic, Gorjan, & Fefferman, Bill. «On Quantum Obfuscation». arXiv:1602.01771 [quant-ph], febrero de 2016. arXiv.org, http://arxiv.org/abs/1602.01771.

[30] Christiano, Paul. «The Informed Oversight Problem». Medium, 4 de julio de 2017, https://ai-alignment.com/the-informed-oversight-problem-1b51b4f66b35.

[31] Armstrong, Stuart, et al. «Thinking Inside the Box: Controlling and Using an Oracle AI». Minds and Machines, vol. 22, n.o 4, noviembre de 2012, pp. 299-324. DOI.org (Crossref), doi:10.1007/s11023-012-9282-2.

[32] Drexler, K.E. (2019): «Reframing Superintelligence: Comprehensive AI Services as General Intelligence», Technical Report #2019-1, Future of Humanity Institute, University of Oxford .

[33] Nielsen, Michael A., and Isaac L. Chuang. «Quantum computation and quantum information». 10th anniversary ed, Cambridge University Press, 2010.